\documentclass[showpacs, pra, aps, twocolumn,10pt]{revtex4}
\usepackage{mathrsfs}
\usepackage{array}
\usepackage{amsmath}
\usepackage{graphicx}
\usepackage{amstext}
\usepackage{bm}
\usepackage{amsfonts}

\begin{document}
\title{Preparing macroscopic mechanical quantum superpositions via photon detection}
\author{Huiping Zhan}
\affiliation{Department of Physics, Huazhong Normal University, Wuhan 430079, China}
\author{Gaoxiang Li}
\affiliation{Department of Physics, Huazhong Normal University, Wuhan 430079, China}
\author{Huatang Tan}
\email{tht@mail.ccnu.edu.cn}
\affiliation{Department of Physics, Huazhong Normal University, Wuhan 430079, China}


\begin{abstract}
In this paper, we propose a feasible scheme for generating the Schr\"{o}dinger cat states of a macroscopic mechanical resonator in pulsed cavity optomechanics. Starting with cooling the mechanical oscillator to its ground state, a red and a blue pulses with different powers are simultaneously employed to drive the cavity to achieve squeezed mechanical states. Subsequently, a second red pulse is utilized to generate the macroscopic mechanical quantum superpositions, conditioned on the detection of cavity output photons. Finally, after being stored in the resonator for a period of time, the mechanical state is mapped, with a third red pulse, to the cavity output field used for state verification. Our approach is generic and can also be used to produce other kinds of non-Gaussian mechanical states, like optical-catalysis nonclassical states.
\end{abstract}
\maketitle

\emph{\textbf{Introduction}} --- The generation of macroscopic quantum superposition states, such as the Schr\"{o}dinger cat states \cite{cat}, has always attracted a lot of research interests, due to their potential applications not only in intrinsic fundamental tests of quantum physics, like decoherence and quantum-classical boundary \cite{fm1,fm2,fm3,fm4}, but also in various quantum technologies \cite{ap1,ap2,ap3,ap4,ap5}. To date, quantum superpositions have been realized in a variety of
physical systems, e.g., photonic \cite{pho1,pho2,pho3,pho4,pho5,pho6}, atomic or molecular systems \cite{atm1,atm2,atm3}, superconducting quantum interference devices \cite{su1,su2}, ranging from microscopic to mesoscopic systems.

With the rapid development in quantum cavity optomechanics in the past decade, much interest has been focused on generating various quantum states of macroscopic mechanical resonators \cite{opm}. Experiments nowadays have realized mechanical states in cavity optomechanical systems, e.g., mechanical squeezed and/or entangled states \cite{squ1, squ2} and optomechanical entangled state \cite{omen1},  which are Gaussian and achieved in the regime of linear optomechanics merely with strong drive. Non-Gaussian quantum superpositions can exhibit negative Wigner functions, indicating genuine nonclassicality \cite{nwg}. Recent proposals have thus been put forward to unconditionally obtain quantum superpositions of macroscopic mechanical resonators \cite{cat1,cat2, cat4, cat5, cat6, cat7, cat8}. Nevertheless, generating non-Gaussian quantum superposition states in a deterministic way remains a challenge currently in cavity optomechanics because of weak nonlinear optomechanical coupling. To circumvent this, quantum measurements are now seen as a particularly promising route to obtaining non-Gaussian mechanical states \cite{ng1,ng2,ng3,ng4}. It should be noted that conditioned on the detection of a cavity output photon, the heralded generation of single phonon Fock states of a mechanical resonator has been experimentally realized very recently \cite{spn}.

In this Letter, based on the fact that experimental realization of Gaussian mechanical squeezed states and heralded non-Gaussian single phonon states, as mentioned above, we propose a feasible scheme for producing the Schr\"{o}dinger cat states of a macroscopic mechanical oscillator via photon subtraction in pulsed cavity optomechanics. Photon subtraction has already been proven to be an efficient way to produce non-Gaussian optical states \cite{pho2, pho3, pho5, ps1,ps2,ps3,ps4,ps5}.
Further, photon-number-resolving detectors are also readily available to use for the multiphoton-state generation \cite{ps5,pn1,pn2,pn3}. Our scheme can be divided into three steps: we first consider the generation of squeezed states of a mechanical resonator by simultaneously driving an optomechanical cavity with a red and a blue laser pulses. We next focus on the utilization of a second red pulse to generate mechanical cat states, conditioned on the detection (subtraction) of odd or even photons from the cavity. We finally investigate the storage and verification of the generated mechanical states by using a third red pulse to mapping the phononic states to the cavity output subject to homodyne detection for the state verification.

\emph{\textbf{Step 1}}: \emph{Generating mechanical squeezed states} --- We first consider the generation of squeezed states of a mechanical resonator with the method originally proposed by one of us \cite{tan} and experimentally realized in Ref. \cite{squ2}. As shown in Fig.1, the cavity optomechanical setup under our investigation consists of a mechanical resonator coupled to the intracavity field, characterized by the Hamiltonian $\hat H=\omega_c\hat a_c^\dag \hat a_c+\omega_m\hat b_m^\dag \hat b_m+g_0\hat a_c^\dag \hat a_c(\hat b_m^\dag+ \hat b_m)$. Here the annihilation operator $\hat a_c~(\hat b_m)$ denote the cavity (mechanical) mode of resonance $\omega_c$ ($\omega_m$), and $g_0$ represents single-photon OM coupling. At first, we consider that the precooling of the mechanical oscillator is performed by
using a dilution refrigerator operating at a base temperature, e.g., 35~mK with a mechanical resonance $\omega_m/2\pi=5.25$ GHz, as achieved in Ref.\cite{spn}. To generate mechanical squeezed states, we further consider that the cavity is simultaneously driven by a blue- and a red-detuned pulses with the same duration denoted by $\tau_{rb}$. For strong driving, the two drives respectively induce two parametric processes for the quantum fluctuations of the cavity and mechanical modes, i.e., optomechanical parametric downconversion and beam-splitter-like mixing, characterized by
$\hat H_{1b}=g_{1b}(\hat a_c\hat b_m+\hat a_c^\dag \hat b_m^\dag)$ and
$\hat H_{1r}=g_{1r}(\hat a_c\hat b_m^\dag+\hat a_c^\dag \hat b_m)$,
on the condition that $\omega_m \gg\{g_{1b/r},\kappa_c, \gamma_m\}$. The linear collective coupling $g_{1b/r}=g_0\sqrt {n_{1b/r}}$ and $n_{1b/r}= P_{1b/r}\kappa_c/[\hbar \omega_c(\omega_m^2+\kappa_c^2/4)]$, with $P_{1b/r}$ the drive powers. With the linearized Hamiltonian and taking into account the cavity dissipation and mechanical damping, the motion of equations for the variables $\hat a_c$ and $\hat b_m$ can be found to be
\begin{subequations}
\begin{align}
\frac{d}{dt}\hat{a}_c&=-\frac{\kappa_c}{2}\hat a_c-ig_{1b}\hat b_m^\dag-ig_{1r}\hat b_m+\sqrt{\kappa_c}\hat a_c^{\rm in}(t),\\
\frac{d}{dt}\hat{b}_m&=-\frac{\gamma_m}{2}\hat b_m-ig_{1b}\hat a_c^\dag-ig_{1r}\hat a_c+\sqrt{\gamma_m}\hat b_m^{\rm in}(t),
\end{align}
\end{subequations}
where  $\kappa_c$ is the cavity dissipation rate and $\gamma _m$ the mechanical damping rate. The vacuum noise operator $\hat a_c^{\rm in}(t)$ satisfy nonzero correlations $\langle \hat a_c^{\rm in}(t)\hat a_c^{\rm in \dag}(t')\rangle=\delta(t-t')$, the thermal mechanical noise operator $\hat b_m^{\rm in}(t)$ has the nonzero correlations $\langle \hat b_m^{\rm in\dag}(t)\hat b_m^{\rm in}(t')\rangle=\bar {n}_{\text{\rm th}}\delta(t-t')$ and $\langle \hat b_m^{\rm in}(t)\hat b_m^{\rm in\dag}(t')\rangle=(\bar{n}_{\rm th}+1)\delta(t-t')$, and $\bar{n}_{\rm th}=(e^{\hbar\omega_m/k_BT}-1)^{-1}$ is the mean thermal phonon number of the environment at temperature $T$, $k_B$ the Boltzmann constant.
\begin{figure}
\centerline{\scalebox{0.30}{\includegraphics{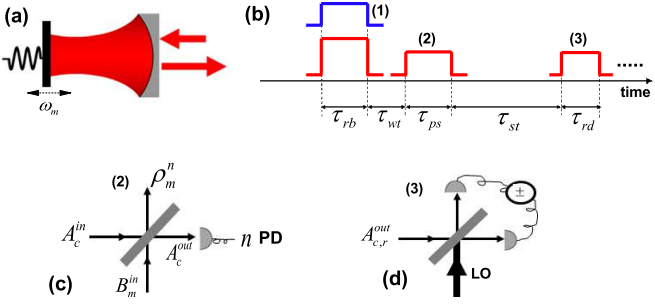}}}
 \caption{(a) The schematic plot of a cavity optomechanical system. (b) The pulse sequence applied in the scheme. A first pair of red and blue pulses of duration $\tau_{rd}$ are used to generate the mechanical squeezed state. After a waiting time $\tau_{wt}$, a second red pulse of duration $\tau_{ps}$ is applied to achieve the mechanical quantum superpositions, conditioned on the detection of $n$ photons from the cavity with a photon-number-resolving detector. As shown in (c), the second red pulse induces the beam-splitter transformation between the cavity input $\hat A_{c}^{\rm in}$ (in vacuum or squeezed states) and the mechanical squeezed input $\hat B_{m}^{\rm in}$, and the detection $n$ photons (PD) at the cavity output $\hat A_c^{\rm out}$ heralds the desirable mechanical state $\hat \rho_{m}^n$. After the storage of the achieved mechanical state in the mechanical resonator for a time $\tau_{st}$, a third red pulse with duration $\tau_{rd}$ is again used to mapping the phononic state to the cavity output field $\hat A_{c,r}^{\rm out}$ which is under homodyne detection for state tomography, as depicted in (d) where ``LO'' denotes local oscillator.}
\end{figure}
It can be easily found that in the steady-state regime, the mechanical oscillator is prepared in a squeezed thermal state
\begin{align}
\rho_m^{\rm ss}=\hat S_m(\xi_m)\Big(\sum_{m=0}^\infty \frac{{\bar m}^m}{(1+\bar m)^{m+1}}|m\rangle\langle m|\Big)\hat S_m^\dag(\xi_m),
\label{rho1}
\end{align}
where the squeezing operator $\hat S_m(\xi_m)=\exp(-\frac{r_m}{2}\hat b_m e^{-i\varphi_m}+\frac{r_m}{2}\hat b_m^\dag e^{i\varphi_m})$, with the squeezing degree $r_m=\tan^{-1}(g_{1b}/g_{1r})$ and the squeezing angle $\varphi_m=\pi$ for simplicity. The mean phonon number of the thermal state $\bar m=\frac{\gamma_m\bar {n}_{\rm th}[4g^2+\kappa_c\gamma_m(\kappa_c+\gamma_m)]}{(\kappa_c+\gamma_m)(4g^2+\kappa_c\gamma_m)}$, with $g=\sqrt{g_{1r}^2-g_{1b}^2}$. We can see that for negligible effective damping rate ($\gamma_m\bar {n}_{\rm th}\approx0$), we have $\bar m=0$, the mechanical resonator can be driven, by the cavity dissipation, into a squeezed vacuum in the steady-state regime, that is the steady mechanical state $|\psi\rangle_m^{\rm ss}=\hat S_m(\xi_m)|0_m\rangle$. The squeezing degree $r_m$ is only dependent on the ratio $g_{1b}/g_{1r}$. Note that because the mean phonon number $\langle\hat b_m^\dag \hat b_m\rangle\sim e^{r_m}$, the large value of $r_m$ leads to $\langle\hat b_m^\dag \hat b_m\rangle\gg |\beta_m^{\rm ss}|^2$ ($\beta_m^{\rm ss}$ being the classical mechanical amplitude) and brings about the linearization of the above Hamiltonian invalid. In addition, it should be noted that the minimum time for achieving the steady mechanical squeezed state is about $\kappa_c^{-1}$ and therefore the pulse duration should be chosen as $\kappa_c^{-1}\ll \tau_{rb}\ll \gamma_m ^{-1}$.

\emph{\textbf{Step 2}}: \emph{Subtracting phonons from the mechanical squeezed states} --- After obtaining the mechanical squeezed state, we proceed to consider the generation of mechanical quantum superpositions via photon detection. Once the mechanical resonator is prepared in the squeezed states with the time $\tau_{rb}$, the driving is then turned off. At this time, the cavity field is also in a squeezed thermal state for finite mechanical damping rate. We in principle need to wait for some time to return the cavity to vacuum, that is $\kappa_c^{-1}\ll \tau_{wt}\ll\gamma_m^{-1}$, during which the prepared mechanical state almost remains. After that time, a second red pulse with duration $\tau_{ps}$ and power $P_{ps}$ is applied to drive the cavity again. Such driving also yields the liner mixing
$\hat H_{2r}=g_{2r}(\hat a_c\hat b_m^\dag+\hat a_c^\dag\hat b_m)$, for $\omega_m \gg\{g_{2r},\kappa_c, \gamma_m\}$. The equations of motion for the operators $\hat a_c$ and $\hat b_m$ are
$\frac{d}{dt}\hat{a}_c=-\frac{\kappa_c}{2}\hat a_c-ig_{2r}\hat b_m^\dag+\sqrt{\kappa_c}\hat a_c^{in}$ and $\frac{d}{dt}\hat{b}_m=-\frac{\gamma_m}{2}\hat b_m-ig_{2r}\hat a_c+\sqrt{\gamma_m}\hat b_m^{in}$.
In the bad-cavity limit, i.e., the cavity dissipation rate $\kappa_c\gg \{g_{2r},\gamma_m\}$, on the time scale $\kappa_c^{-1}\ll\tau_{ps} \ll \gamma_m^{-1}$ the cavity mode is almost in its steady states and can thus be adiabatically eliminated from the evolution of the system. Meanwhile, the mechanical damping can be neglected during this time. Exactly for this reason, the waiting time is unnecessary, i.e., $\tau_{wt}=0$, since the initial squeezed cavity state, generated at the end of the first step, is ruined by the cavity dissipation in the steady-state regime. By defining the normalized temporal modes of the input and output light of the cavity \cite{hof}:
$\hat A_c^{\rm in}=\sqrt{\frac{2G_{2r}}{e^{2G_{2r}\tau_{ps}}-1}} \int_{0}^{\tau_{ps}}dte^{G_{2r}t}\hat a_c^{\rm in}(t)$, and
$\hat A_c^{\rm out}=\sqrt{\frac{2G_{2r}}{1-e^{-2G_{2r}\tau_{ps}}}} \int_{0}^{\tau_{ps}}dte^{-G_{2r}t}\hat a_c^{\rm out}(t)$,
where $G_{2r}=4g_{2r}^2/\kappa_c$ and the cavity output $\hat a_c^{\rm out}=\sqrt{\kappa_c}\hat a_c+\hat a_{\rm in}$, one can obtain the following input-output relation
\begin{subequations}
\begin{align}
\hat A_c^{\rm out}&=-\mathcal{T}\hat A_c^{\rm in}+i\sqrt{1-\mathcal{T}^2}\hat B_m^{\rm in},\\
\hat B_m^{\rm out}&=-i\sqrt{1-\mathcal{T}^2}\hat A_c^{\rm in}+\mathcal {T}\hat B_m^{\rm in}.
\end{align}
\label{mbs}
\end{subequations}
Here we have defined the mechanical input and output modes $\hat B_m^{\rm in}=\hat b_m(0)$ and $\hat B_m^{\rm out}=\hat b_m(\tau_{ps})$ in time domain, and $\mathcal T=e^{-G_{2r}\tau_{ps}}$. Evidently, Eq.(\ref{mbs}) describes an effective beam-splitter-like transformation between the output and input mechanical and cavity modes. Therefore, the output modes $\hat A_c^{\rm out}$ and $\hat B_m^{\rm out}$ can be considered as the output fields from a beam splitter, with the input fields $\hat A_c^{\rm in}$ and $\hat B_m^{\rm in}$ ``injected'' into the beam-spitter.  The transmissivity $\mathcal T$ of the effective beam splitter can be adjusted by changing the pulse duration $\tau_{ps}$. When a photon-number-resolving detector detects $n$ photons from the optical output $\hat A_c^{\rm out}$, it implies the subtraction of $ n$ phonons from the mechanical input state of $\hat B_m^{\rm in}$, for the vacuum input of $\hat A_c^{\rm in}$. We therefore realize $n$-phonon subtraction operation, allowing us to achieve non-Gaussian mechanical states.

The mechanical input state $\hat B_m^{\rm in}$ is just the mechanical squeezed state generated in the last step, i.e., $\hat \rho_m^{\rm ss}$, and for a generic consideration we assume that the optical input mode $\hat A_c^{\rm in}$ is also in a squeezed vacuum state $|\psi\rangle_c =\hat S(\xi_c)|0_c\rangle$ for $\xi_c=r_ce^{i\varphi_c}$, with the squeezing angle $\varphi_c=0$. This can in principle be complemented by injecting broadband squeezed light (as a reservoir) into the cavity directly.  The density matrix $\hat\rho_{\rm out}$ of the two modes $\hat A_c^{\rm out}$ and $\hat B_m^{\rm out}$ from the effective beam splitter can be expressed as
\begin{align}
\hat\rho_{\rm out}=\frac{1}{\pi^2}\int\chi_{\rm out}(\eta,\zeta)\hat D_m^{\dag}(\eta)\hat D_c^{\dag}(\zeta)d^2\eta d^2\zeta,
\end{align}
in terms of the corresponding Wigner characteristic function $\chi_{\rm out}(\eta,\zeta)=\exp (-\frac{1}{2}x\sigma x^{T})$ of the output-field state, where the vector
$x=(\eta_r,\eta_i,\zeta_r,\zeta_i)$ and $\hat D_{c/m}$ are the displacement operators corresponding to the output fields $\hat A_c^{\rm out}$ and $\hat B_m^{\rm out}$. The nonzero elements of correlation matrix $\sigma$ are
$\sigma_{11}=A_m\mathcal T^2+A_c\mathcal R^2$,$\sigma_{22}=B_m\mathcal T^2+B_c\mathcal R^2$, $\sigma_{33}=A_m\mathcal R^2+A_c\mathcal T^2$, $\sigma_{44}=B_m\mathcal R^2+B_c\mathcal T^2$, $\sigma_{13}=\mathcal T\mathcal R(A_m-A_c)$, $\sigma_{24}=\mathcal T\mathcal R(B_m-B_c)$, with $\mathcal R=\sqrt{1-\mathcal T^2}$ and $A_y/B_y=2\langle \hat o_y^\dag\hat o_y\rangle\mp2\langle\hat o_y^2\rangle+1$ $(o=\{a, b\}, y=\{c,m\})$.

When detecting $n$ photons for the output field $\hat A_c^{\rm out}$, the conditional mechanical state of the mechanical mode $\hat B_m^{\rm out}$ is then obtained as
\begin{align}
\hat\rho_{m}^{n}=\mathcal {N}_{n}\langle n|\hat\rho_{\rm out}|n\rangle,
\end{align}
with the normalization factor $\mathcal N_{n}=1/\langle n|Tr_m[\hat\rho_{\rm out}]|n\rangle$.
The probability $P_r\equiv \mathcal {N}_{n}^{-1}$ for successfully detecting $n$ photons is given by
\begin{align}
P_r&=Tr_m\langle n|\hat\rho_{\rm out}|n\rangle\nonumber\\
&=\frac{1}{\pi}\sum_{k=0}^{n}\frac{(-1)^kC_n^k}{k!}\Big[\sum_{l=0}^{k}C_k^lF_1^{-\frac{1}{2}-l}F_2^{\frac{1}{2}- k+l}\nonumber\\
&~~~~~~~~~~~\times \Gamma(\frac{1}{2}+l)\Gamma(\frac{1}{2}+k-l)\Big],
\end{align}
where $\Gamma(x)$ is the Gamma Function and $F_j=(\sigma_{j+2,j+2}+1)/2$. The characteristic function $\chi_{m}(\lambda)\equiv Tr[\hat\rho_{m}^{n}\hat D_m(\lambda)]$ for the conditional state $\hat\rho_m^{n}$ can be found to be
\begin{align}
\chi_m(\lambda)&=\frac{\mathcal N_n}{\pi}e^{-\frac{1}{2}(\sigma_{11}-\frac{\sigma_{13}^2}{2F_1})\lambda_r^2-\frac{1}{2}(\sigma_{22}-\frac{\sigma_{24}^2}{2F_2})\lambda_i^2}
\sum_{k=0}^n\frac{(-1)^kC_n^k}{k!}\nonumber\\
&~~\times\sum_{l=0}^{k}\Big[C_k^lF_1^{-\frac{1}{2}-l}F_2^{-\frac{1}{2}- k+l}\Gamma(\frac{1}{2}+l)\Gamma(\frac{1}{2}+k-l)\nonumber\\
&~~\times K(-l,\frac{1}{2},-\frac{(\sigma_{13}\lambda_r)^2}{4F_1})K(l-k,\frac{1}{2},-\frac{(\sigma_{24}\lambda_i)^2}{4F_2})\Big],
\end{align}
where $K(x,y,z)$ is the Kummer function. Then,
the Wigner function of the conditional mechanical state $\hat\rho_m^{n}$ can be obtained by performing the Fourier transform on the characteristic function:
\begin{align}
W_m(\beta_m)=\frac{1}{\pi^2}\int\chi_m(\lambda)e^{\beta_m\lambda^*-\beta_m^*\lambda}d^2\lambda.
\label{wg}
\end{align}
The fidelity $F_{\pm}$ for the mechanical state $\hat\rho_m^{n}$ to the cat's states
$|\psi_{\alpha_m}\rangle_{\pm}=(|\alpha_m\rangle\pm|-\alpha_m\rangle)/\sqrt{1\pm \exp [-2\alpha_m^2]}$ can be calculated via
\begin{align}
F_{\pm}=\frac{1}{\pi}\int\chi_m(\lambda)\chi_{\pm}(\lambda)d^2\lambda,
\end{align}
where $\chi_{\pm}(\lambda)$ are the characteristic functions of the cat states $|\psi\rangle_{\pm}$.
The nonclassicality of the mechanical superpositions can be measured by the negativity of the wigner function
\begin{align}
N_m=\int |W_m(\beta_m)|d^2\beta_m-1.
\end{align}

\begin{figure}[t]
\centerline{\scalebox{0.35}{\includegraphics{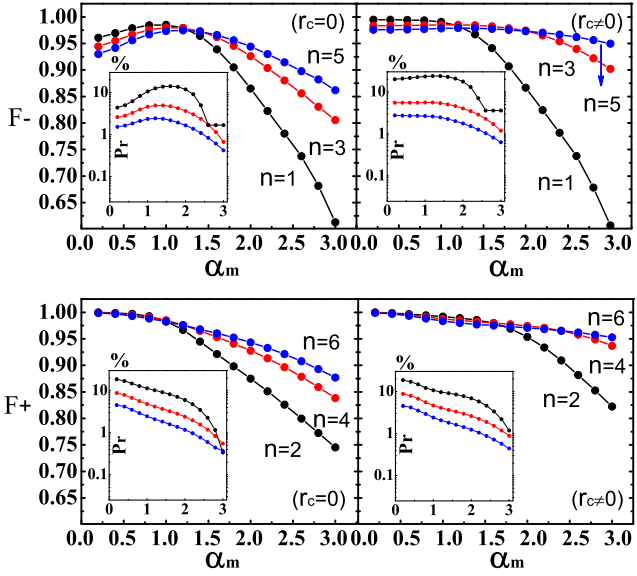}}}
\caption{The fidelities $F_\pm$ between the generated mechanical superposition $\hat \rho_m^n$ and the cat state $|\psi_{\alpha_{m}}\rangle$ for different number $n$ of detected photons. The fidelities $F_\pm$ are optimized with respect to the effective transmissivity $\mathcal T$ and the cavity squeezing $r_c$, with the mechanical squeezing $r_m=1.1$. The inserts plot the corresponding probability $P_r$ of the photon detection.}
 \label{f4}
\end{figure}

In Fig.2, we plot the fidelities $F_\pm$ for the cases that the cavity injected squeezing $r_c=0$ and $r_c\neq0$. The fidelities are optimized with respect to the squeezing parameters $r_c$ and the effective transmissivity $\mathcal T$.  We consider the parameters in the experiment \cite{spn}, i.e., $\omega_m/2\pi=5.25$ GHz, $g_0/2\pi=869$ kHz, $\kappa_c/2\pi=846$ MHz, $\gamma_m/2\pi=13.8$ kHz, and $ T=35$ mK. We thus have the mechanical damping time $\gamma_m^{-1}\approx11~\mu$s. The mechanical squeezing is $r_m=1.1$, by setting the powers of the red and blue pulses $P_{1r}\approx 80~\mu$W and $P_{1b}\approx 50~\mu$W. We see from the figure that for the odd-photon detection and $r_c=0$, when the cat's amplitude $\alpha_m$ exceeds the small value $\alpha_m\approx 1.2$, the fidelity $F_-$ increases with the increase of the number $n$ of detected photons, but with the decease of the probability $P_r$. At this amplitude, the fidelity is maximal, i.e., $F_-\approx 0.98$, with the almost same negativity $N_m\approx 0.41$, and independent of $n$. Once the amplitude $\alpha_m$ exceeds this value, the fidelity drops quickly as $\alpha_m$ increases. For example, we have the fidelity $F_-\approx \{ 0.63, 0.81,0.86\}$ at $\alpha_m=3$ for $n=\{1, 3,5\}$, with the negativity $N_m\approx\{0.41,0.49,0.53\}$. While for the even-photon detection and $r_c=0$, the fidelity $F_+$ also increases as $n$ rises but it monotonously drops as $\alpha_m$ grows up. We have the fidelity $F_+\approx \{ 0.75, 0.84,0.88\}$ at $\alpha_m=3$ for $n=\{2, 4, 6\}$, with the negativity $N_m\approx\{0.44,0.51,0.53\}$. When the cavity squeezing input is present ($r_c\neq0$), the fidelities $F_{\pm}$ are obviously improved, compared to that for $r_c=0$, except for $n=1$. For example, for $n=3$ the fidelity is enhanced from $F_-\approx0.81$ to $0.9$ at $\alpha_m=3$ and the negativity $N_m$ is also improved to 0.64. Fig.3 depicts the density plots of the Wigner function $W_m(\beta_m)$ of the generated mechanical states for $r_c=0$. We see that the obvious interference between the two cat components can be found, except for the case of $n=1$ although with the high fidelity $F_-\approx 0.98$ with respect to $\alpha_m=1.2$. Here, the probabilities for achieving the mechanical states in (a)-(f) are respectively $Pr=\{12.6\%, 3.66\%,0.42\%, 10.2\%, 2.39\%, 0.37\%\}$.
\begin{figure}[t]
\centerline{\scalebox{0.27}{\includegraphics{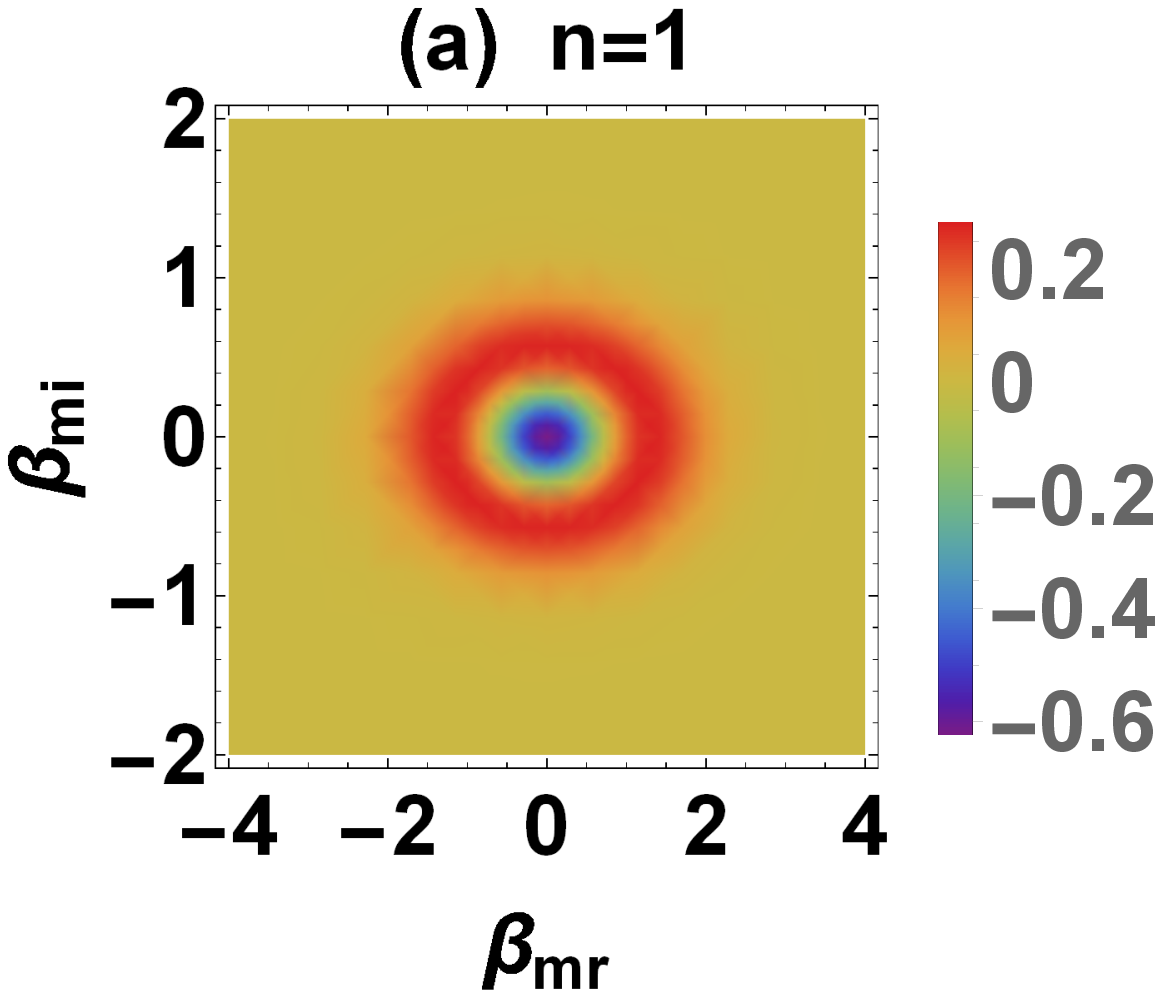}{\includegraphics{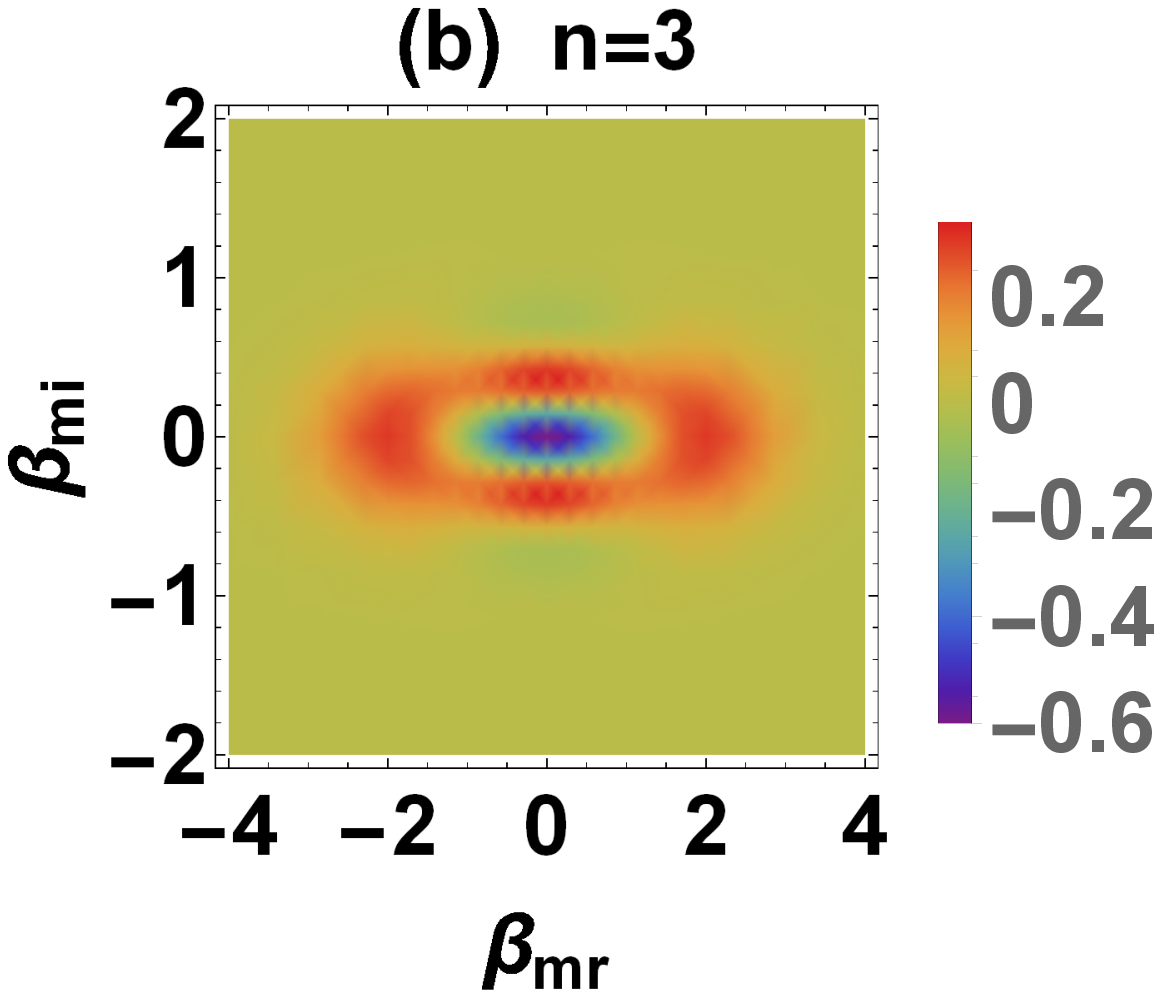}{\includegraphics{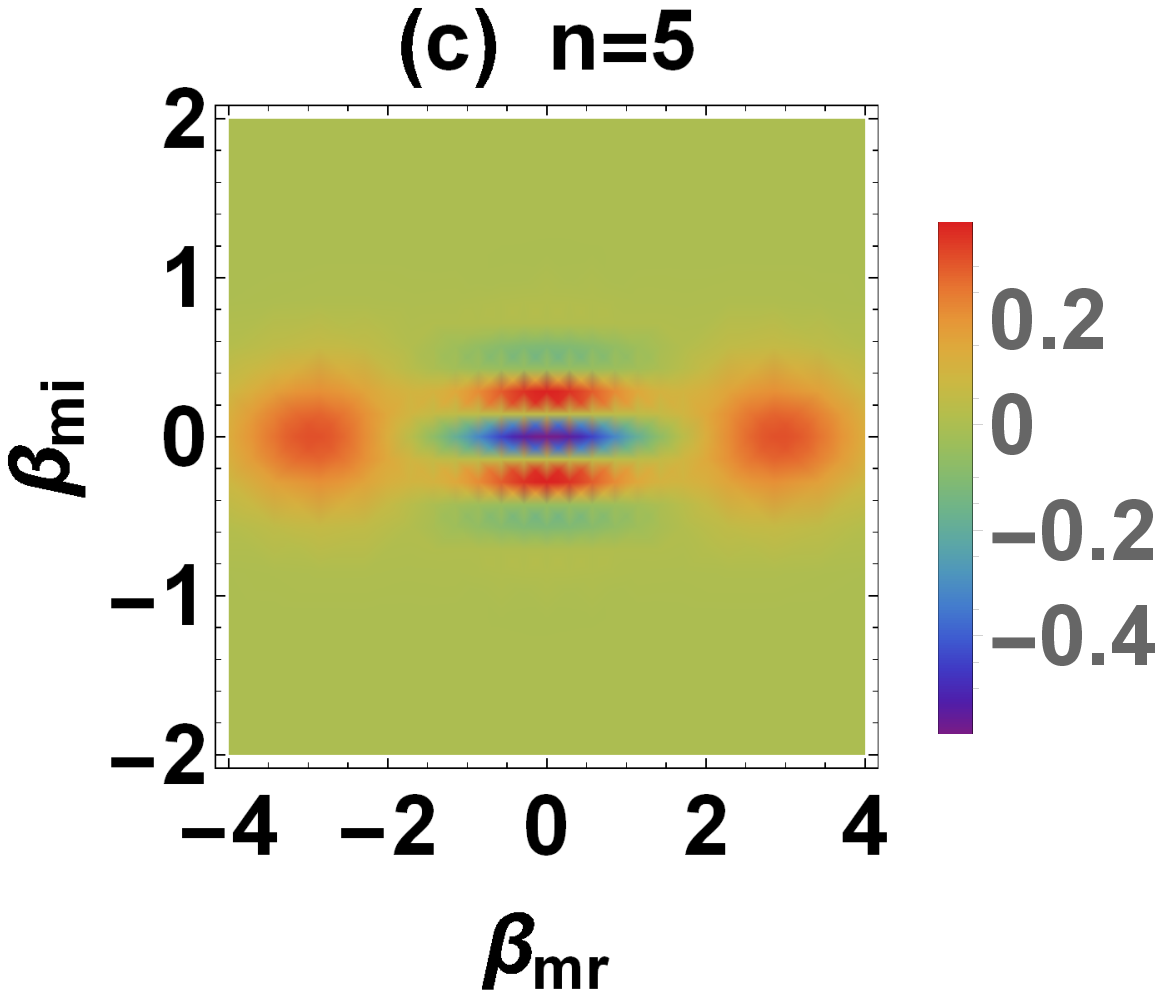}}}}}
\centerline{\scalebox{0.27}{\includegraphics{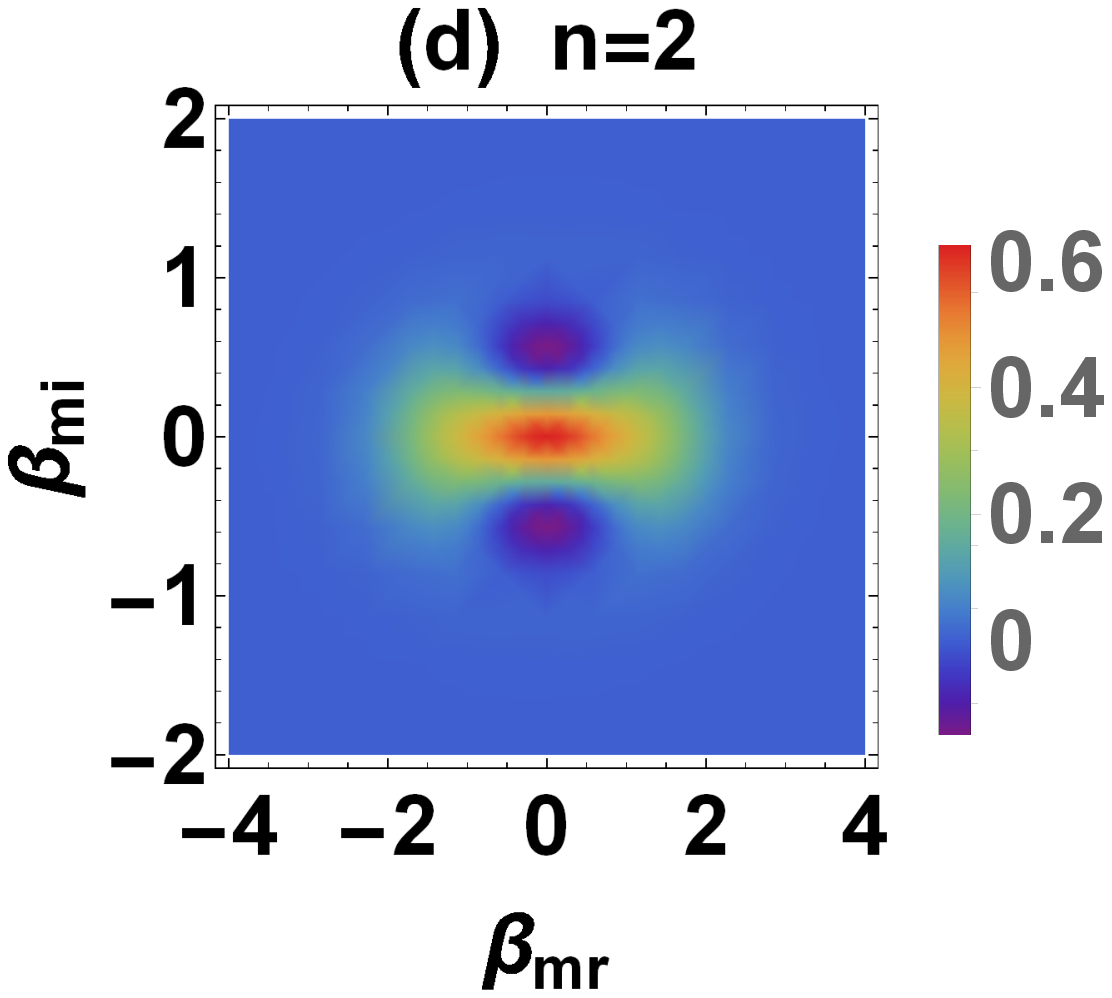}~~~{\includegraphics{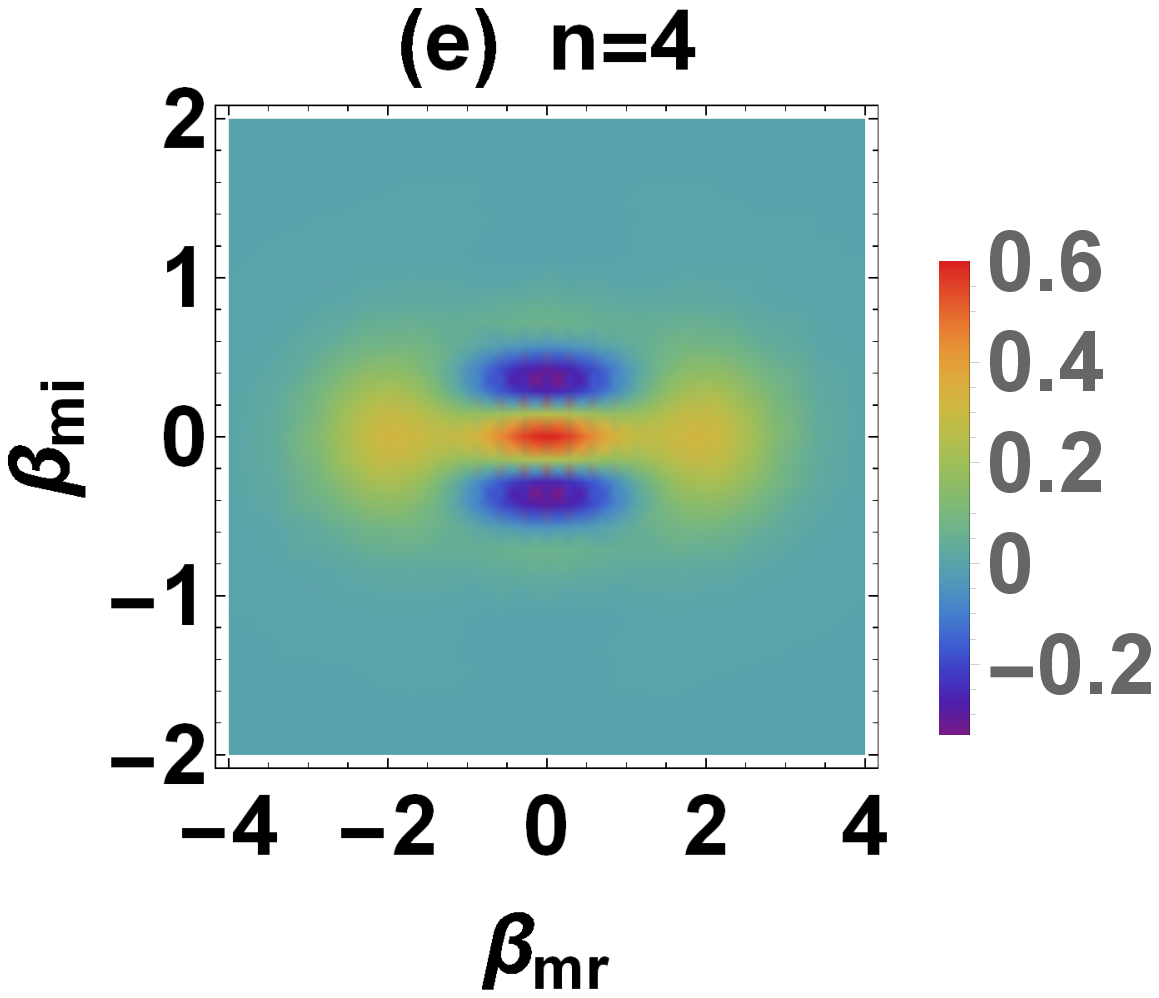}~~~{\includegraphics{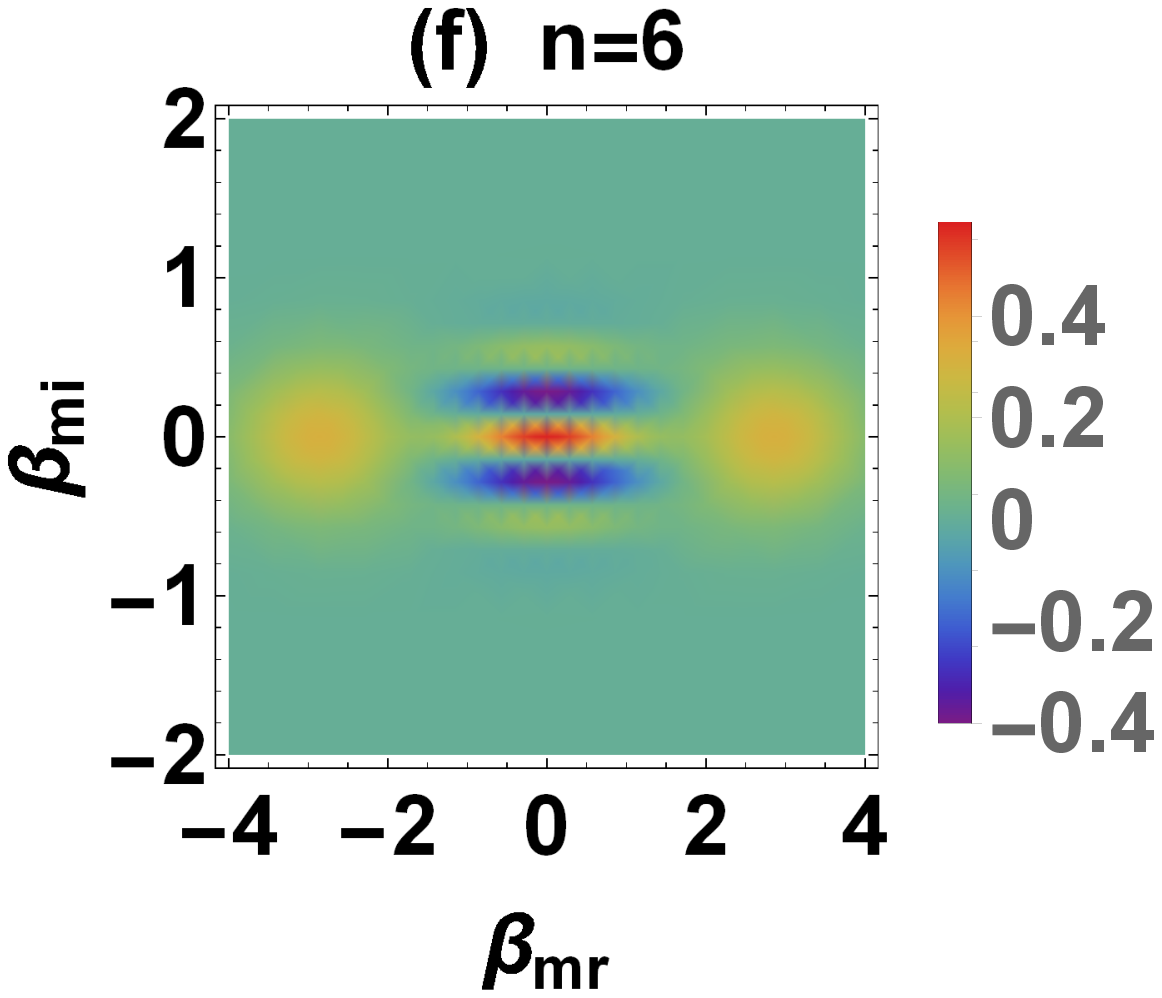}}}}}
 \caption{The density plots of the Wigner function $W_m(\beta_m)$ of the mechanical states shown in Fig.2 for $r_c=0$. (a), (b), and (c) depict the mechanical states generated with $\mathcal T=\{0.51,0.65,0.77\}$, whose fidelity $F_-=\{0.98, 0.93,0.86\}$ to the corresponding cat states with amplitudes $\alpha_m=\{1.2,2,3\}$, and (d), (e), and (f) are the mechanical states generated with $\mathcal T=\{0.46,0.59,0.70\}$, for the fidelity $F_+=\{0.97,0.93,0.88\}$ to $\alpha_m=\{1.2,2,3\}$.}
 \label{f2}
\end{figure}

\emph{\textbf{Step 3}}: \emph{Storing and verifying the mechanical superpositions} --- Once the non-Gaussian mechanical superpositions are generated, we hope to store them in the high-quality mechanical resonator for some time $\tau_{st}$ and then verifying them by mapping the phononic states to a cavity output pulse subject to verification detection via applying again a red pulse to drive the cavity and induce optomechanical state exchange. During the storage period, the mechanical state is under a free evolution in the mechanical environment and its characteristic function $\chi_m(\lambda)$ satisfies the following motion of equation
\begin{align}
\frac{\partial}{\partial t}\chi_m(\lambda,t)=-\gamma_m(\hat A_\lambda+(2\bar n_{\rm th}+1){|\lambda|}^2)\chi_m(\lambda,t),
\label{eqs}
\end{align}
where the operator $\hat A=\lambda\frac{\partial}{\partial\lambda}+\lambda^*\frac{\partial}{\partial\lambda^*}$. The time-dependent solution of Eq.(\ref{eqs}) can be obtained with the operator-ordering theorem \cite{barn}, explicitly given by
\begin{align}
\chi_m(\lambda,t)=e^{(\bar n_{\rm th}+\frac{1}{2})(e^{-\gamma_m t}-1){|\lambda|}^2}\chi_m(\lambda e^{-\frac{\gamma_m t}{2}},0),
\end{align}
for the initial state $\chi_m(\lambda,0)=\chi_m(\lambda)$.

\begin{figure}[t]
\centerline{\scalebox{0.32}{\includegraphics{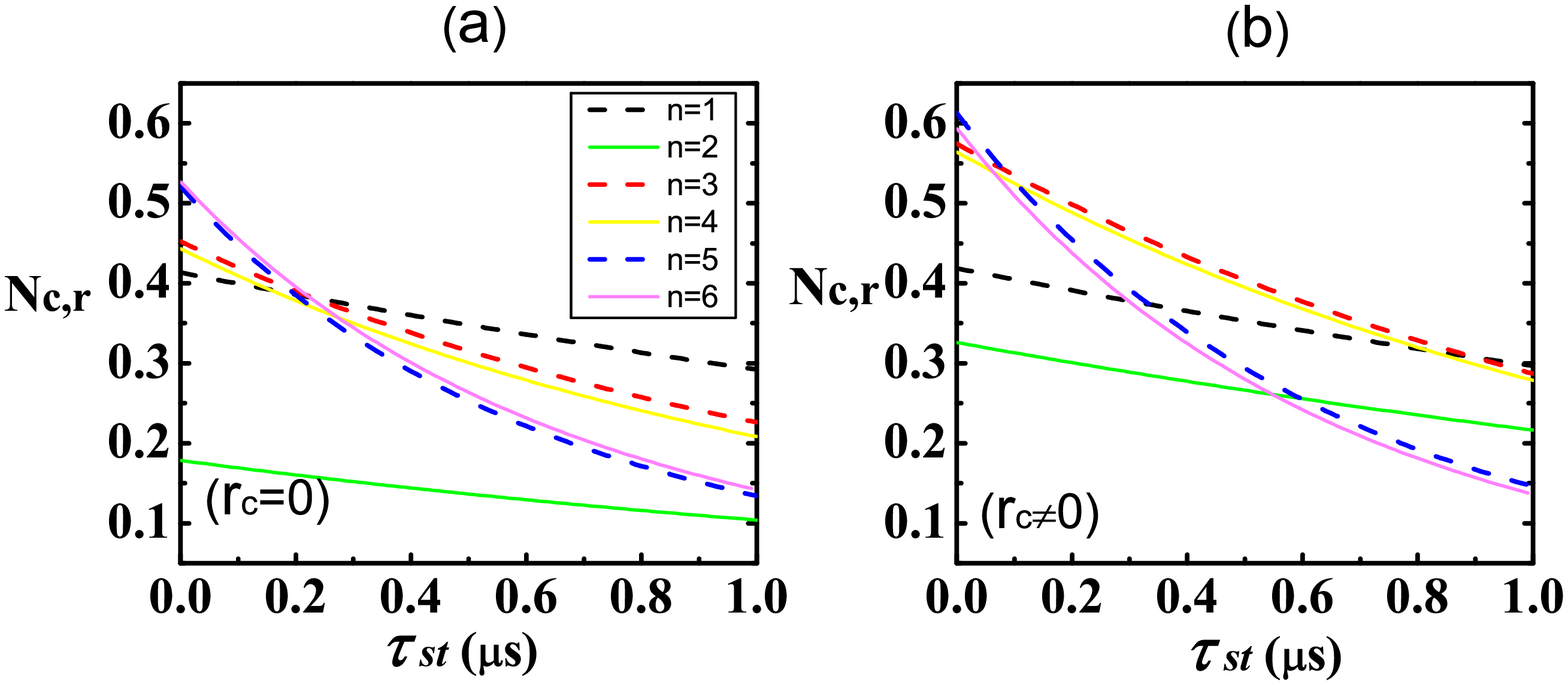}}}
\centerline{\scalebox{0.32}{\includegraphics{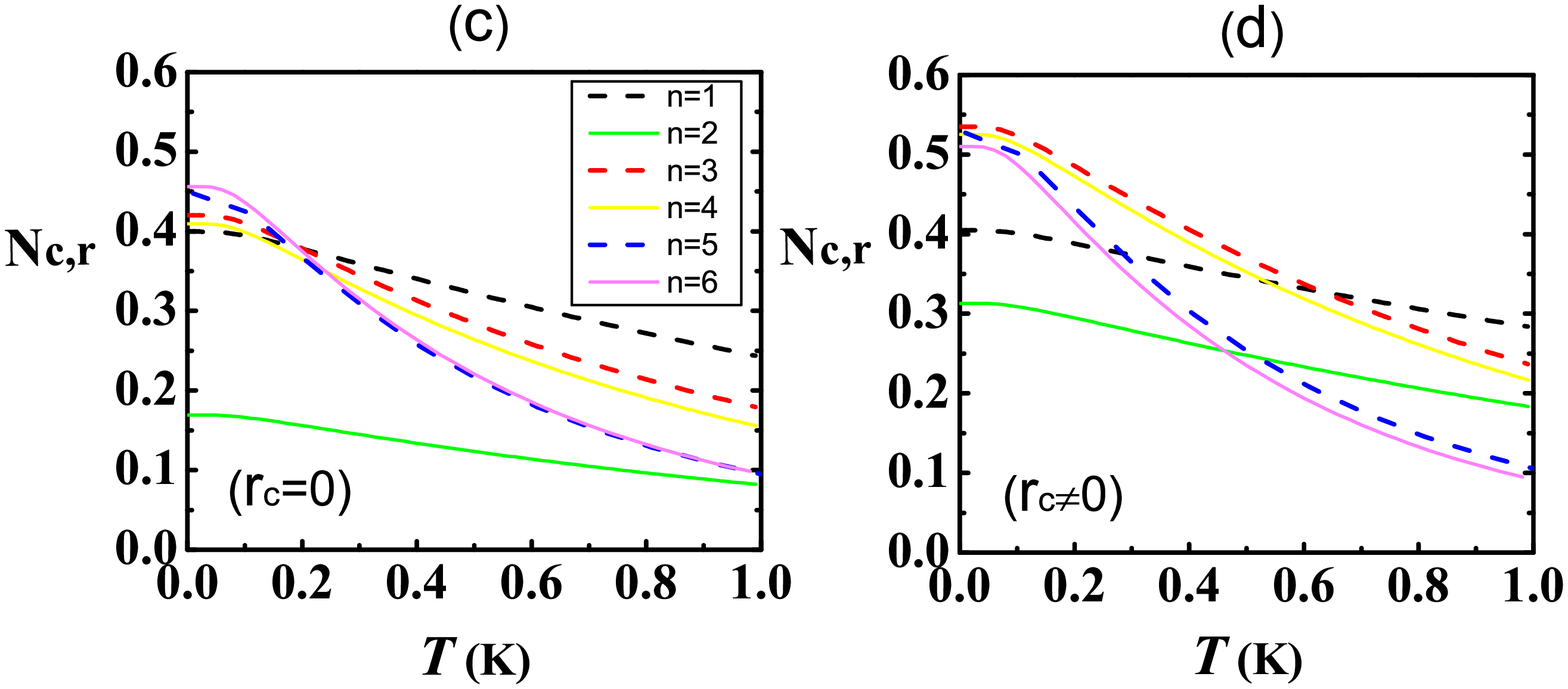}}}
\caption{The dependence of the negativity $N_{c,r}$ of the readout cavity output field on the storage time $\tau_{st}$ and temperature $T$ for the mechanical states shown in Fig.3. In (a) and (b), the temperature $T=35$~mK, and in (c) and (d) the storage time $\tau_{st}=100$~ns.}
\end{figure}
After a storage period $\tau_{st}$, we consider the optical readout of the stored state $\chi_m(\lambda,\tau_{st})$. To this end, a red laser pulse of duration $\tau_{rd}$ and power $P_{3r}$ is applied again to read out the mechanical states via the interaction $\hat H_{3r}=g_{3r}(\hat a_c\hat b_m^\dag+\hat a_c^\dag\hat b_m)$. Likewise, when the pulse duration $\tau_{rd}$ should satisfy $\kappa_c^{-1}\ll\tau_{rd}\ll \gamma_m ^{-1}$ and $\kappa_c\gg g_{3r}$, the input-output relation of the cavity and mechanical modes is the same as Eq.(\ref{mbs}). It is clearly shown that the cavity output field $\hat {A}_{c,r}^{out}\simeq i\hat {B}_{m,r}^{in} $ for the ideal case that $e^{-G_{3r}\tau_{rd}}\to 0$, with $G_{3r}=4g_{3r}^2/\kappa_c$, implying that the mechanical state is completely mapped to the cavity output field. For the nonzero value of the factor $e^{-G_{3r}\tau_{rd}}$, the characteristic function of the cavity output state after the mapping can be derived as
\begin{align}
\chi_{c,r}^{\rm out}(\zeta)=\chi_m\big(\sqrt{1-e^{-2G_{3r}\tau_{rd}}}\zeta,\tau_{rd}\big)e^{-\frac{1}{2}|ie^{-G_{3r}\tau_{rd}}\zeta|^2},
\end{align}
from which the corresponding Wigner function $W_{c,r}(\beta)$ and negativity $N_{c,r}$ can be calculated.

In Fig.4, the dependence of the negativity $N_{c,r}$ of the cavity output states on the storage time $\tau_{st}$ and temperature $T$ are plotted for the generated mechanical states shown in Fig.3. Here we choose the drive power $P_{3r}\approx 150~\mu$W, the coupling $g_{3r}/2\pi\approx 65~$MHz, and the pulse duration $\tau_{rd}= 30~$ns. It is shown that even for the storage time $\tau_{st}=1~\mu s$, much shorter than the mechanical damping time $\gamma_m^{-1}\approx11~\mu s$, the negativity, as the indicator of nonclassicality, can still be achieved. When considering the pulse durations $\tau_{rb}=30$~ns and $\tau_{ps}=30~$ ns (e.g. with the power $P_{2r}\approx0.2~\mu$W, yielding $\mathcal T=0.99$), the condition of the whole time of the protocol $\tau_{rb}+\tau_{ps}+\tau_{st}+\tau_{rd}\ll \gamma_m^{-1}$ is ensured. In addition, as shown in Fig.4 (c) and (d), when the temperature $T=1$~K (i.e., $\bar n_{\rm th}=3.5$ for which the mechanical damping time $\gamma_m\bar n_{\rm th}\approx3~\mu$s), the mechanical quantum superpositions with the negative Wigner functions can be achieved. Note that the temperature also affects the mechanical squeezing generated in the first step.

\textbf{\emph{Conclusion}} ---
We propose in this paper a feasible scheme for the generation, storage, and verification of the Schr\"{o}dinger cat states of a macroscopic mechanical resonator in pulsed cavity optomechanics. After preparing the mechanical oscillator in a squeezed mechanical state, we consider the utilization of a red driving pulse to achieve the macroscopic mechanical quantum superpositions, which is conditioned on the detection of odd or even photons from the cavity. The cavity squeezed input can effectively enhance the fidelity of the mechanical superpositions. We finally consider the storage  of the generated mechanical states in the mechanical resonator for a period of time and then the readout for verification by mapping the phononic states to the cavity output field. The fidelity and nonclassicality of the mechanical states are studied in detail, and the effect of thermal noise is also evaluated. The present scheme can also be used to produce other kinds of non-Gaussian mechanical states, in analogy to linear optics with nonclassical inputs. For example, one can use it to prepare mechanical superposition states via photon catalysis \cite{pc1,pc2}. A further investigation will also include the generation of non-Gaussian entangled mechanical states, such as NOON state.

\emph{\textbf{Note added}} --- After completion of this work, we became
aware of a similar preprint paper, arXiv:1909.10624v1 by
I. Shomroni, L. Qiu, and T. J. Kippenberg, on the conditional generation of mechanical quantum superpositions by photon detection.

\section*{Acknowledgment}
This work is supported by the National Natural Science Foundation of China (No.11674120) and the Fundamental Research Funds for the Central Universities (No. CCNU18TS033).
\end{document}